# 'Multi-Nested Pendula': a new concept for vibration isolation and its application to gravitational wave detectors

O. D. Aguiar[1] and M. Constancio Jr.[1]

[1]*Divisão de Astrofísica, Instituto Nacional de Pesquisas Espaciais, Avenida dos Astronautas 1758, 12227-010, São José dos Campos, Brazil*

Adequate vibration isolation is of great importance for the design of any sensitive experiment measuring mechanical motions. The so-called multistage or multipole mechanical low pass filter is a common artifact used for the construction of highly effective vibration isolation. However, the problem with it is the vertical clearance needed inside the vacuum chambers for the construction of numerous, long pendulum stages necessary for effective horizontal vibration isolation. The purpose of this work is to introduce a new concept, called the "multi-nested pendulum" concept, which solves this problem. The attenuation performance of an ideal multistage *nested pendulum* filter is better than that of an ideal multistage common equal pendulum filter by a factor of $N^N$, where N is the number of stages used, making this idea of a *multi-nested pendulum* a very interesting one. The initial results (of stability and system resonances) measured for a prototype reinforce the viability of the practical implementation of such idea. A detailed study for the realization of a cryogenic and fully operational version to be used in advanced LIGO (aLIGO) is going to be done in a future study.

## I. INTRODUCTION

Any experiment that needs precise measurement of motion in mechanical systems must take into account the possibility of data contamination of the data due to unwanted external sources of vibration motion.[1] Manmade sources of vibration can be minimized by careful choice of the site (far from human activities) and control of the environment near the laboratory. Also the choice of a seismically quiet underground site can mitigate *seismic noise*. However, for very sensitive experiments (such as gravitational wave detectors) these preventive measures may not be enough. Thus it becomes clear that adequate vibration isolation is of great importance when designing any sensitive experiment measuring mechanical motions.

The so called multistage or multipole[1] mechanical low pass filter is a common artifact used for the construction of highly effective passive vibration isolation. There are many proposed designs, such as "six-degree-of-freedom" vibration isolators[1], stacks[2], and Taber vibration isolators.[3] For ground-based interferometers, multistage pendulum systems are very effective for vibration isolation[4]. The problem with them is the vertical clearance needed inside the vacuum chambers for the construction of numerous long pendulum stages, which are necessary for effective horizontal vibration isolation. Then, alternative solutions as the "inverted pendulum" concept or the clever Roberts linkage[5] are used as pre-isolator stage[6] in order to compensate the difficulty in using as many as possible long enough pendulum stages for effective very low frequency horizontal vibration isolation.

The purpose of this work is to introduce a new concept, called the 'multi-nested pendulum' concept, which can solve this problem. By putting all pendula 'nested' inside each other we save vertical space and, at the same time, permit the construction of the mechanical filter with as many stages as required. Now the vertical clearance needed is the length of a single pendulum, the common pendulum length of all pendula in the system.

The idea of nested pendula for use in vibration isolation is not new. A two-stage nested pendulum (*nested double pendulum suspension system*[7]) has already been proposed in the literature for vibration isolation systems. This is used, for example, at the two-stage BSC-ISI system for the Advanced Laser Interferometer Gravitational Observatory (aLIGO). In Abbott et al[8] this nested two stage active isolation platform can be seen. The idea of a multi-nested pendulum, which was independently conceived, goes further, which is to have a large number of pendula inside each other, as many as necessary to optimize the attenuation, all with their lengths equal to the vertical clearance available. This multi-nested pendulum idea can be implemented if we use concentric nested cylindrical shells all positioned with their axis vertically oriented. The vertical wall of each cylindrical shell is used to shift up, again and again, the next pendula point of suspension. In principle, with this artifice we can use an "infinite" number of concentric nested cylindrical shells. However, for a given volume available there must be always an ideal number of stages that maximize the attenuation.[9]

## II. THE MULTI-NESTED PENDULA CONCEPT

The multi-nested pendula concept is the natural solution to the simple question: how can one pack into a finite vertical clearance as many pendula as one wants? The solution is pictorially explained in figure 1 for a particular case of five-stage nested pendula.

Each cylindrical shell hung from its bottom flange forms a pendulum and each one is connected to the top flange of the next external one. These five "nested" cylindrical shells or pendula make up the mechanical vibration isolation filter. The cylindrical shells, especially their flanges, should be made from a high mechanical Q material, such as Al5056, in order to minimize thermal noise. The space inside the innermost cylindrical shell can be used for the payload (a circular lateral opening on the walls of all cylindrical shells can be provided for optical access to the payload). The payload would be connected to the innermost cylindrical shell. For clarity, the space between the walls of successive cylindrical shells was exaggerated. However, in practical designs this space can be as small as a couple of inches. Therefore, many stages can be assembled forming a very compact and effective mechanical filter. Furthermore, the flanges can have parallel cuts to their circumferences, forming bending blades, in order to also provide vertical attenuation. These bending blades can be machined in such a way as to operate horizontally when the system is assembled with its nominal load, in order to avoid cross coupling between the horizontal and vertical motions. The cylindrical shell walls can be kept thick enough to avoid low frequency resonances. Additional high frequency mechanical filters placed above the payload can filter these, as well as the wire resonant frequencies, out. The architecture of these nested cylindrical shells is also suitable for cryogenic designs, because the shells can easily provide the necessary low temperature enclosure if one of them is heat sunk to a cold liquid reservoir. This problem will be addressed in a later study.

A possible format of such cylindrical shells is shown in figure 2. Evidently they have to have different sizes in order to fit inside each other. They also can have openings in their walls to allow the performance of optical experiments, such as laser interferometers. Instead of machining the flanges curved, which is difficult to do, the flanges could be machined flat and later bent. Evidently, in this case, heat treatment would be necessary to restore the correct elastic properties to the bent flange. Another way would be to replace the elastic flanges with gas springs[10]. The latter would be more suitable if one needs to fine adjust their spring constants (servo-controlling their pressure) in order to avoid tilting instabilities in the nested system.

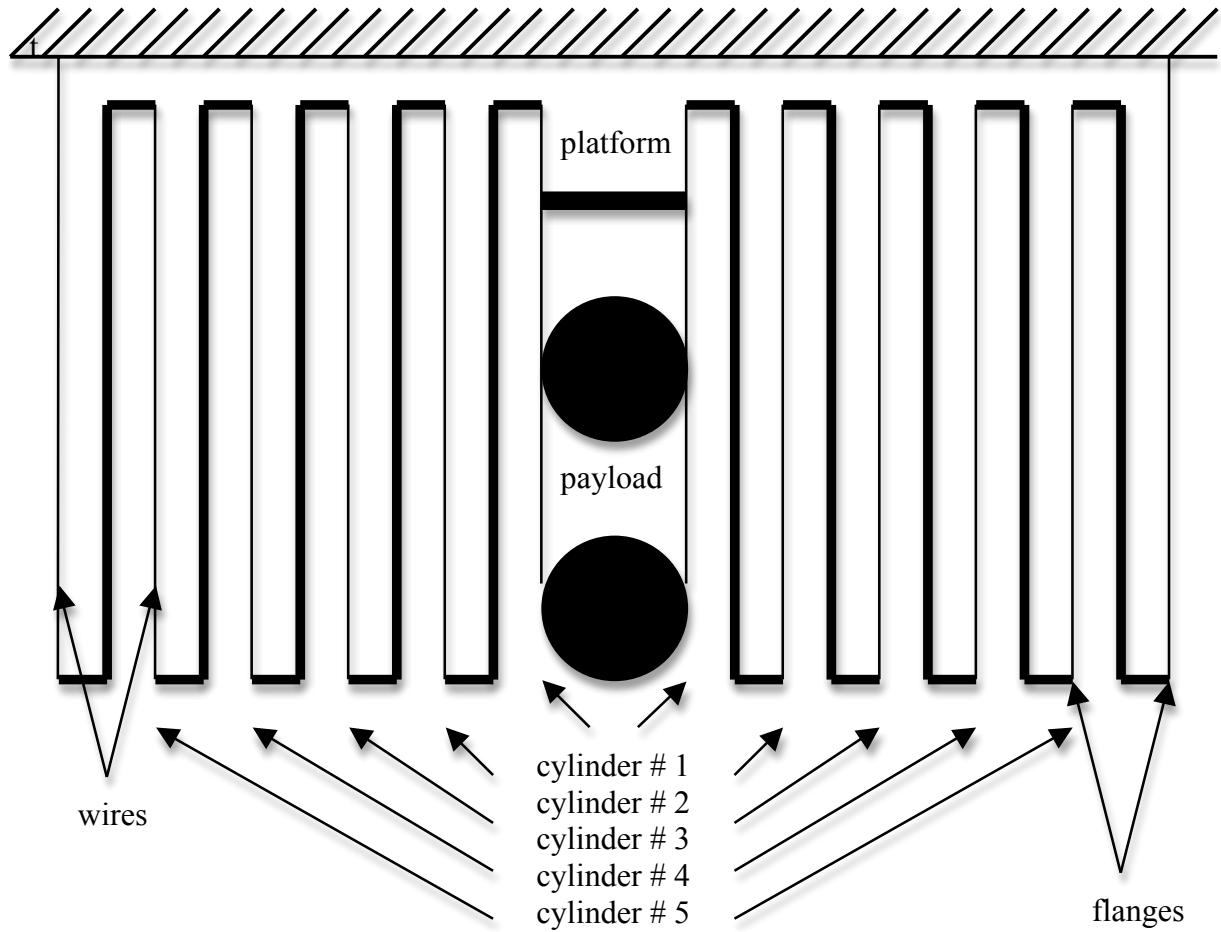

FIG. 1. A schematic cut view of a five-stage nested pendulum is shown. Each cylindrical shell hung from its bottom flange forms a pendulum and each one is connected to the top flange of the neighboring external one. So, in this figure, five "nested" cylindrical shells or pendula make up the mechanical vibration isolation filter. The space inside the most internal cylindrical shell can be used for the payload (a circular lateral opening on the walls of all cylindrical shells can be provided for optical access to the payload). The payload is connected to the innermost cylindrical shell. For clarity, the space between the walls of successive cylindrical shells was exaggerated. However, in actual designs this space can be as small as a couple of inches. Therefore, many stages can be assembled forming a very compact and effective mechanical filter. Furthermore, the flanges can have parallel cuts to their circumferences, forming bending blades, in order to also provide vertical attenuation. These bending blades can be machined or bent in such a way as to operate horizontally when the system is assembled with its nominal load, in order to avoid cross coupling between the horizontal and vertical motions. The cylindrical shell walls can be kept thick enough to avoid low frequency resonances. Additional high frequency mechanical filters placed above the payload can filter these out, as well as the wire resonant frequencies. The architecture of these nested cylindrical shells is also suitable for cryogenic designs, because the shells can easily provide the necessary low temperature enclosure if one of them is heat sunk to a cold liquid reservoir.

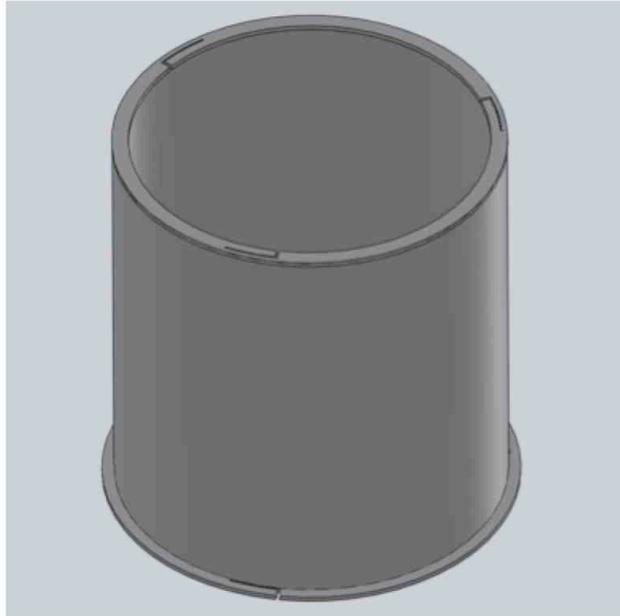

FIG. 2. This shows a possible format of such cylindrical shells (this type is the one used in the prototype). Evidently they have to have different sizes in order to fit inside each other. They also can have openings in their walls to allow the performance of optical experiments, such as laser interferometers. Instead of machining the flanges curved (in the vertical direction), which is difficult to do, the flanges could be machined flat (as they are shown in the figure) and later bent. Evidently, in this case, a heat treatment would be necessary to restore the correct elastic properties to the bent flange. Another way would be to replace the elastic flanges with gas springs. The latter ones would be more suitable if one needs to fine adjust their spring constants (servo-controlling their pressure) in order to avoid tilting instabilities in the nested system.

## III. THEORETICAL COMPARISON BETWEEN THE MULTI-NESTED PENDULA SYSTEM AND THE TRADITIONAL MULTI-CASCADE PENDULA

Pictorial representations of multi-cascade and multi-nested systems can be seen in figure 3.

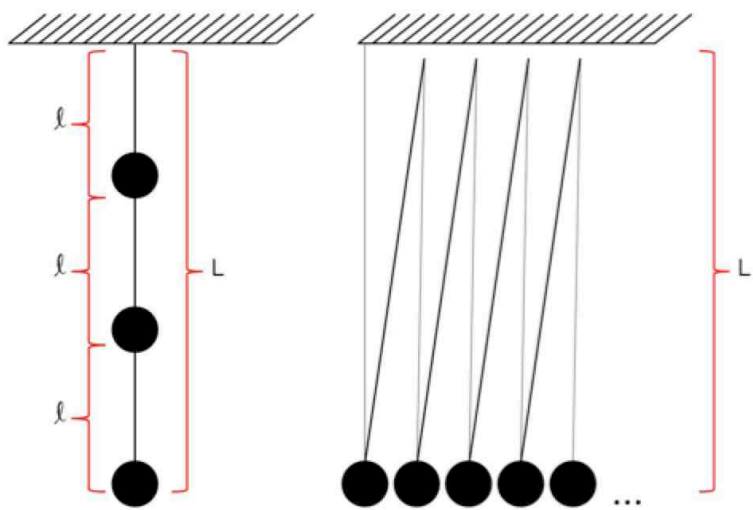

FIG. 3. Pictorial representations of multi-cascade (left) and multi-nested (right) systems are shown. The total length of the cascade system is equal to the common length of each pendulum in the nested system.

Let $\ell$ be the length of each pendulum in the cascade set and L the common length of each pendulum in the nested system. The transmission through the cascade system with N stages is given by

$$T = \left(\frac{\omega_0}{\omega}\right)^{2N} \quad (1)$$

where $\omega_0$ is the resonance frequency of a single isolate stage of the cascade set and $\omega \gg \omega_0$.

If L is equal the total length of the cascade set, L = N $\ell$.

So we can write the resonance frequency of a single stage of the nested set $\omega'_0$ and compare it with the resonance frequency of the single stage cascade system. The result is shown in (2):

$$\omega'_0 = \sqrt{\frac{g}{L}} = \sqrt{\frac{g}{N l}} = \frac{1}{\sqrt{N}}\sqrt{\frac{g}{l}} = \frac{\omega_0}{\sqrt{N}} \quad (2)$$

Now, the transmission through the nested system (T') compared to the transmission through the cascade system (T) is:

$$T' = \left(\frac{\omega'_0}{\omega}\right)^{2N} = \left(\frac{\omega_0}{\omega\sqrt{N}}\right)^{2N} = \left(\frac{1}{N^N}\right)\left(\frac{\omega_0}{\omega}\right)^{2N} = \left(\frac{1}{N^N}\right) T \quad (3)$$

Therefore, for systems occupying the same vertical clearance and for $\omega \gg \omega_0 > \omega'_0$, the transmission through the multi-nested system is $N^N$ better than the transmission through the multi-cascade system.

## IV. AN EXAMPLE: A PROTOTYPE

A full size prototype has been constructed in order to verify the feasibility to implement such multi-nested pendula idea. Pictures of this prototype can be seen in figure 4.

The prototype is composed of five nested cylindrical shells (the number of stages in the system was chosen arbitrarily to be five) made of Al5052 aluminum alloy. The vertical wall thickness of all cylindrical shells is 6.35 mm (1/4") and their horizontal flanges have a thickness of 5/8" (~15.87 mm). The wires/rods are made of stainless (304) steel 3/16" thick (~ 4.76 mm) and 131,3 cm long. Stainless steel pieces were welded to their ends. This can be seen in figure 5. The bottom one is a conical piece that fits in the bottom flange and the top one is a one-inch diameter stainless steel bolt. Rotating a threaded nut around each of these bolts, one can level with precision one cylindrical shell inside another. The 5 cylindrical shells have the following diameters: 121 cm, 108 cm, 96 cm, 84 cm, and 72 cm. This set of 5 cylindrical shells, including the flanges, has a total mass of about 400 kg (the largest outermost one has a mass of 100 kg), which makes it interesting for application to gravitational wave detectors. In particular, a set composed of some of these cylindrical shells could fit inside the present advanced LIGO (aLIGO) vacuum chamber and around its quadruple pendulum. The outermost cylindrical shell could be connected to the active platform and the innermost cylindrical shell to the quadruple pendulum, introducing, in this manner, additional stages of vibration isolation without significantly changing the present system.

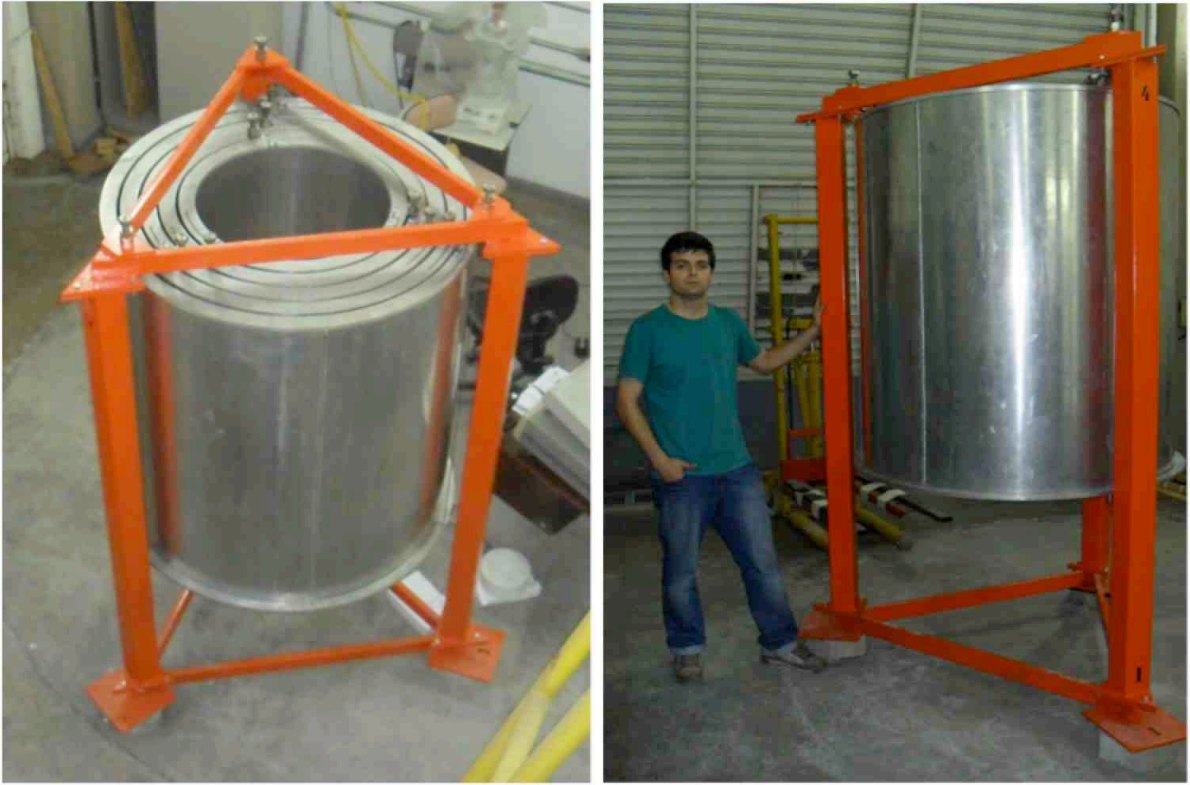

FIG. 4. Pictures of the top (left) and side (right) views of the multi-nested pendula system prototype can be seen.

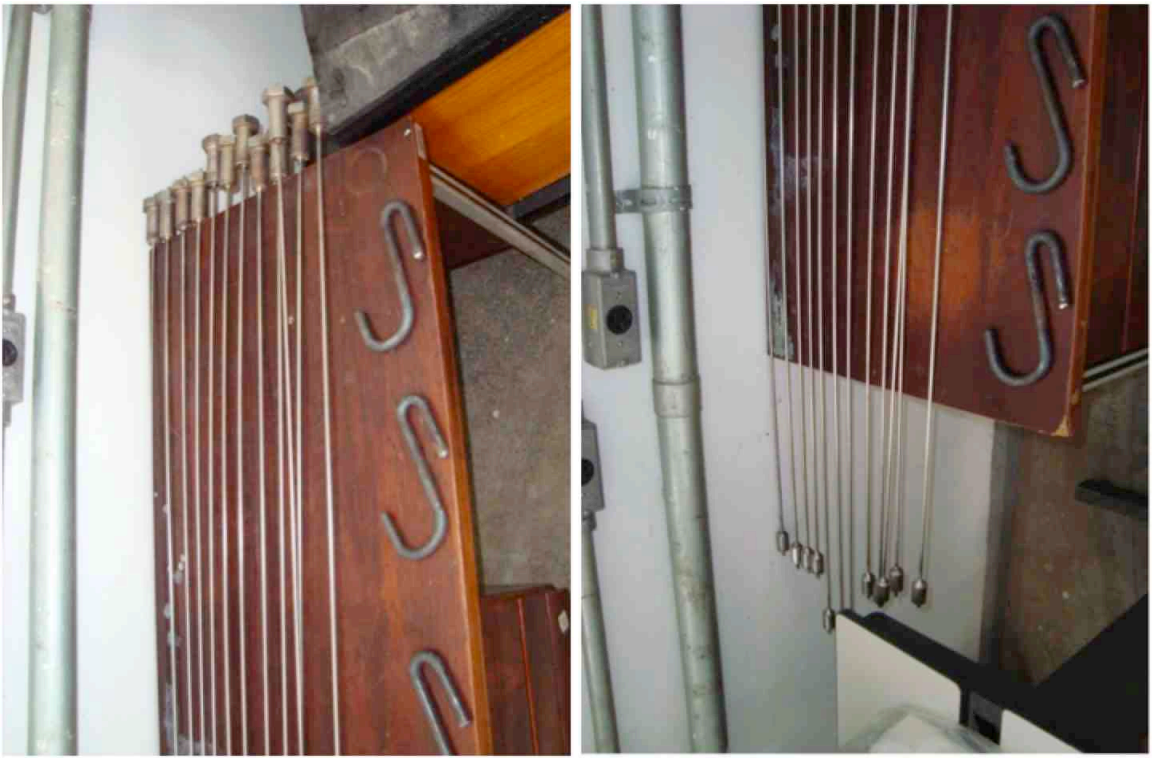

FIG. 5. Pictures of the top (left) and bottom (right) end of the stainless steel rods can be seen.

The system has to be assembled starting from the cylindrical shell with the smallest diameter, because the top flange of the outer cylindrical shell would collide with the bottom flange of the inner cylindrical shell when one try to assemble it from above. After all cylindrical shells are nested the system is then connected to a tall and strong (orange) steel frame (figure 6).

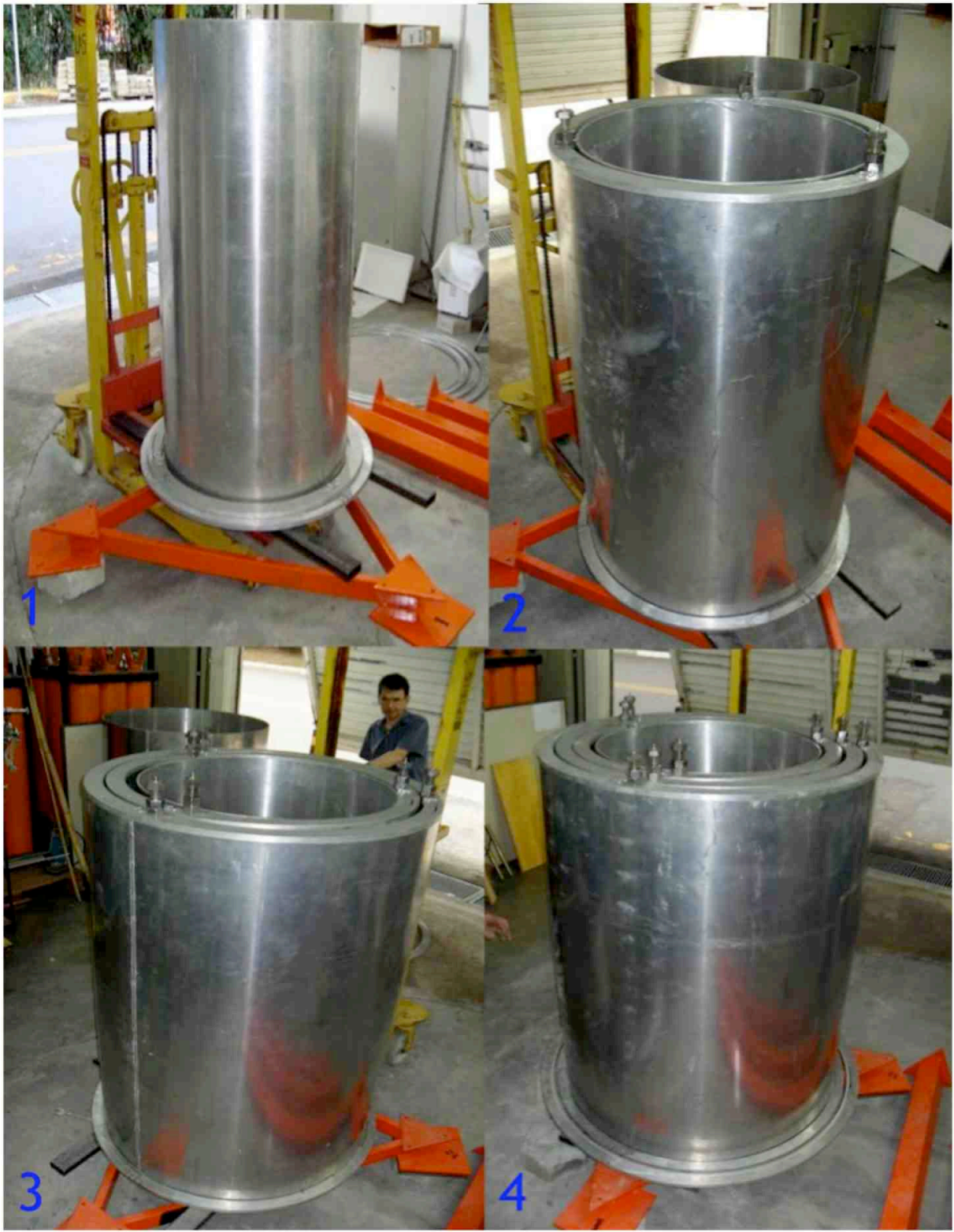

FIG. 6. The sequence for the assembling the multi-nested pendula system can be seen.

**V. PROTOTYPE STABILITY**

After assembling the cylindrical shells, by connecting the stainless steel rods directly to the hardest part of the flanges (not using the flange's long arm springs), the system displayed very good stability. The clearance between the flanges was ~1 cm and all the pendula could oscillate individually, although their oscillations were evidently coupled. It is shown in this paper the results for this assembled configuration. In a later study we intend to explore the assembly by connecting the stainless rods to the flange's long arm springs. Our feeling is that the system will no longer be stable, because small differences in the spring constants of these arms may exist and cause large angle of tilting. If this hypothesis is confirmed in the future, we intend to replace these long arm springs with gas springs. Adjusting the pressure inside the gas springs (with a servo system), we should be able to suppress these spring constant differences and make the system stable again. However, this is a major change in the hardware, which we plan to do in a future work. For the time being, we intend to show the results of a multi-nested pendulum system whose isolation is much worse for the vertical than for the horizontal oscillations.

We have analyzed both theoretically and experimentally (using our prototype) the stability of the whole assembly for tilting supposing we can adjust the vertical springs to equal values (in our present case the vertical springs are sufficiently stiff, so the relative tilting between neighboring cylindrical shells is small. Our conclusion is that, in the present case, it is always stable. When ones tilts the outermost cylindrical shell, all other internal cylindrical shells tilt the same angle, maintaining their axis perpendicular to each other all the time, but their axis are not degenerated (at the same centered position) anymore. All cylindrical shells shift sideways until each one touches its neighboring cylindrical shell. The maximum tilt angle allowed before this happens is arctangent of d/L, where d is the horizontal clearance between the cylindrical shells flanges and L is the common length of the pendula. Therefore, the outermost cylindrical shell needs to have one of its sides lifted by D times (d/L), where D is its diameter, in order to have any of the internal cylindrical shells touching each other. If, for example, the horizontal clearance between successive flanges d is 1 cm, D is 115 cm, and L is 130 cm, the maximum tilt allowed for the most external cylindrical shell is when one of its sides is lifted by approximately 9 mm. Evidently, this can be avoided with the adjusting bolt-nut system at the top of the rods.

**VI. SYSTEM RESONANCES**

The success of such a proposed mechanical filter design depends on one important question: do the system resonant frequencies fall within the low frequency band of the *nested pendulum* filter? If so, the whole project is compromised. In the case of advanced LIGO (aLIGO) the target vibration isolation bandwidth to be significantly improved is in the range ~2-10 Hz, so the system should not have any resonances there.

The pendulum frequencies are all below this range, as they should in order to cause vibration isolation in the frequency range immediately above (~2-10 Hz). Their measured frequencies, using piezoelectric sensors, were 0.236, 0.519, 0.793, 1.040, 1.310 Hz for the five x and y degenerated horizontal translational modes and 0.310, 0.580, 0.862, 1.144, and 1.461 Hz for the five horizontal rotational modes (around the z axis). The spectrum of these resonances can be seen in figure 7. The modes related to the vertical oscillations of the cylindrical shells

(up and down in the z direction and tilting around the x or y axis) were not measured, because the long arm flange springs were not used. They will be measured in the future, when we address the instability for tilting with (perhaps) gas springs with servo-controlled pressure.

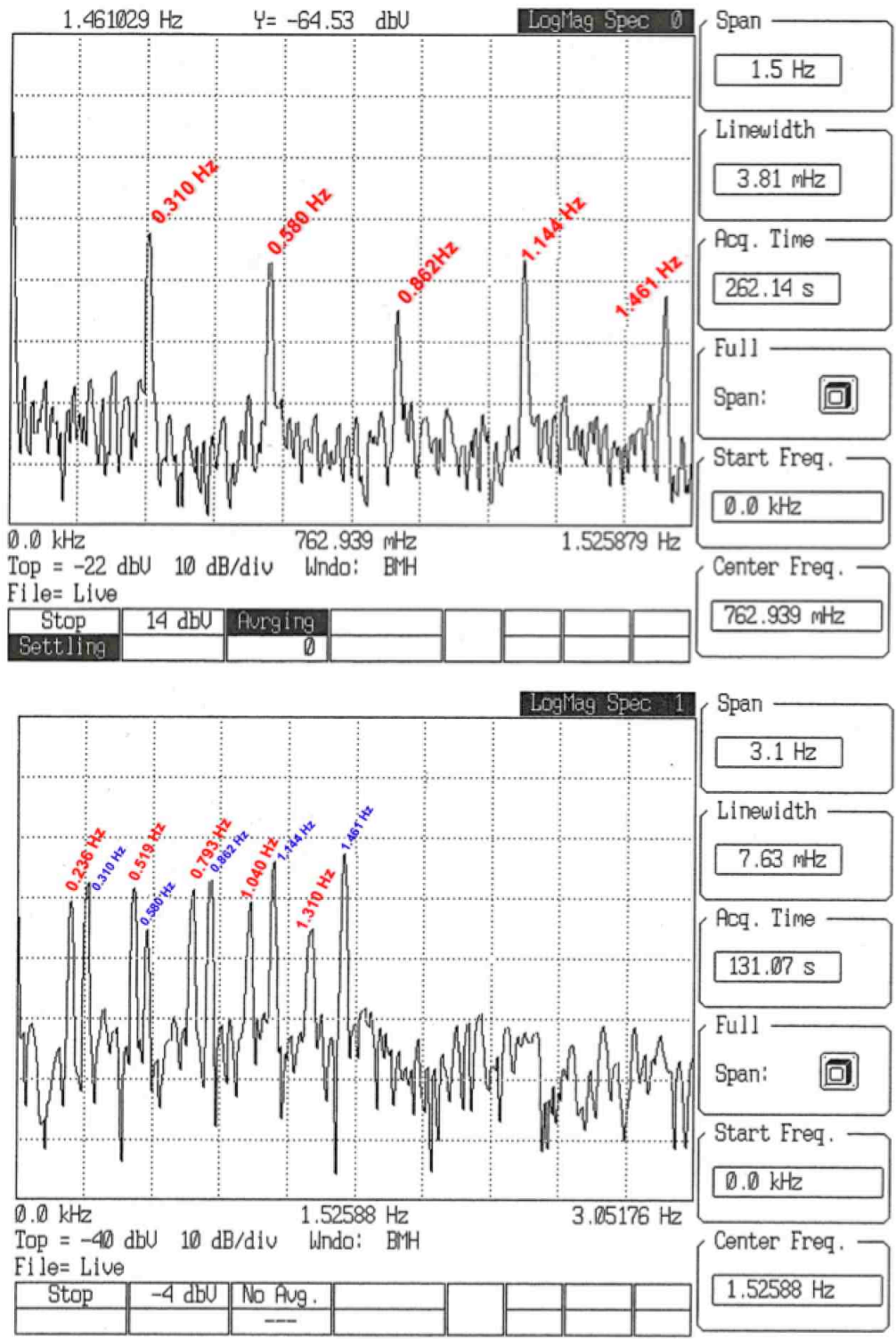

FIG. 7. Spectra showing the five horizontal rotational modes around the z axis (top spectrum) and the five x and y degenerated horizontal translational modes together with the previous modes (bottom spectrum). It is easy to excite the rotational modes alone, but this is not true for the horizontal translational ones.

On the other hand, the measured and predicted transverse resonances of the rod and the measured and predicted resonances of the largest outer cylindrical shell of the prototype are shown in Table 1. All measurements were made with piezoelectric sensors, and their signals were processed in an FFT spectrum analyzer. The predicted values for the cylindrical shells were calculated using the finite element software provided by SolidWorks$^{TM}$ version 2010 (www.solidworks.com), and the predicted values for the rods used this finite element software and an analytical solution given by Blevins[7]. It was not possible to find a correct analytical solution for the cylindrical shells due to the complexity of the set cylindrical shell-flange. Some of the predicted frequencies for the cylindrical shell were not found by the measurements, perhaps because the piezoelectric sensors were placed in some of their mode nodes.

| RESONANCES OF THE LARGEST CYLINDRICAL SHELL (HZ) | | ROD RESONANCES (HZ) | | |
|---|---|---|---|---|
| PREDICTED | MEASURED | PREDICTED | | MEASURED |
| | | Finite Element (violin modes) | BLEVINS (violin modes) | |
| 29.24 (circumferential, i=2, j=0) | 24.66 | | | 6.6 (?) |
| 43.22 (circumferential-axial?) | 39.06 | | | 14.65 (?) |
| 72.53 (circumferential, i=3, j=0) | | | | 22.46 (?) |
| 88.01 (circumferential-axial?) | 84.50 | 37.84 | 35.54 (n=1) | 37.11 |
| 110.29 (circumferential-axial?) | 115.72 | 73.17 | 71.08 (n=2) | 74.71 |
| 127.02 (circumferential-axial, i=4, j=1) | 128.90 | | | 82.52 (?) |
| 142.77 (circumferential-axial?) | | 105.6 | 106.62 (n=3) | 97.17 |
| 148.85 (circumferential-axial, i=5, j=1) | 163.82 | 137.18 | 142.16 (n=4) | 134.77 |
| 182.31 (circumferential-axial, i=6, j=1) | 173.34 | | | 142.58 (?) |
| 221.92 (circumferential-axial, i=3, j=2?) | 220.46 | | | 157.23 (?) |
| 224.18 (circumferential-axial, i=4, j=2?) | 224.36 | 180.04 | 177.69 (n=5) | 166.99 |
| 235.06 (circumferential-axial, i=7, j=1) | | | | |
| 282.04 | 270.07 | 233.06 | 213.23 (n=6) | 220.70 |
| 284.24 (circumferential-axial, i=5, j=2) | 288.57 | 263.74 | 248.78 (n=7) | |
| 292.54 (flange modes) | 292.97 | 294.41 | 284.32 (n=8) | 281.25 |
| 301.20 (torsion-circumferential-axial) | | | | |
| 303.03 (torsion-circumferential-axial) | | THESE ROD RESONANCE DATA ARE FOR A 1400 N TENSION IN THE ROD. AFTER THE ADDITION OF THE INSIDE PAYLOAD, THE TENSION WILL INCREASE TO 4000 N, SO ALL THESE FREQUENCIES WILL GO UP BY ~60%. | | |
| 307.70 (circumferential-axial, i=6, j=2) | | | | |
| 329.15 | 321.29 | | | |
| 331.11 (circumferential-axial, i=7, j=2) | | | | |
| 334.22 | | | | |
| 341.54 | | | | |
| 350.24 | | | | |

TABLE 1. The measured and predicted resonances of the rod and cylindrical shells are shown. All measurements used piezoelectric sensors and a spectrum analyzer. The predicted values for the outermost cylindrical shell were calculated using the finite element software provided by SolidWorks[TM] version 2010 (www.solidworks.com), and the predicted values for the external stainless steel rods (L= 1.413 m and D=0.00476 m) were calculated using the finite element software provided by SolidWorks[TM] and an analytical solution given by Blevins[7]. It was not possible to find a correct analytical solution for the cylindrical shell case due to the complexity of the set cylindrical shell-flange. The measurements with (?) mark may be noise or vertical oscillation resonances, for which the piezoelectric sensor attached to the rod would be responsive, because they were not predicted by the theory or finite element simulation. Note that the vertical oscillations resonances will be moved to much lower frequencies when the flexure arm flanges start to be used.

Many axial and radial-circumferential flexural modes calculated using SolidWorks[TM] (Cosmos) for the single shell of figure 2 (0.605 m radius and 0.00635 m wall thickness) can be seen in figure 8. The first natural frequency is at 29.24 Hz. This and other calculated frequencies are in good agreement with the measurements. The mode shapes of axial and radial-circumferential modes are illustrated in figure 9.

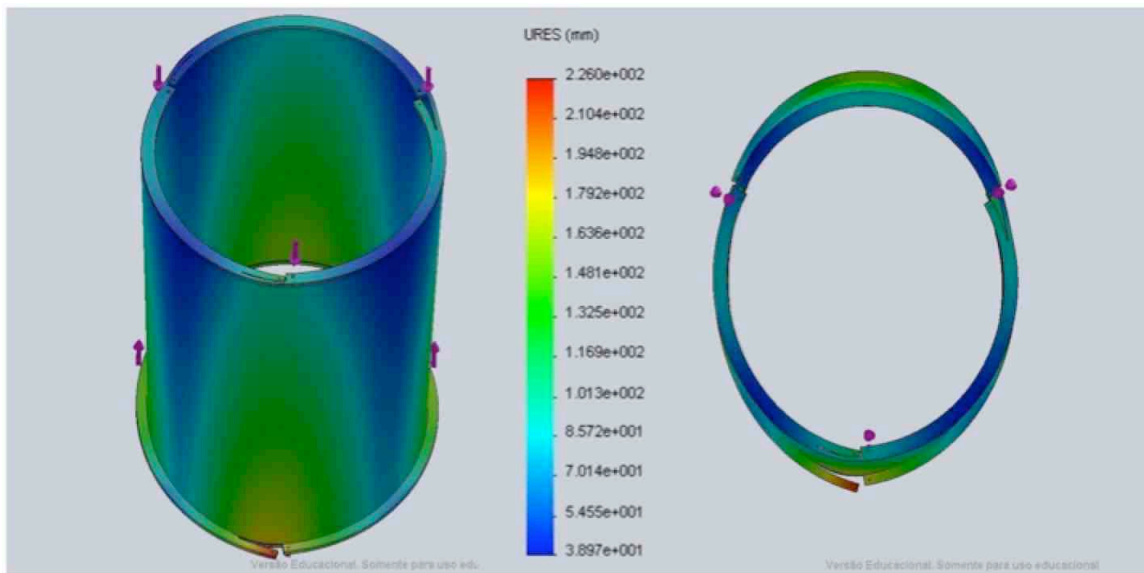

a) 29.24 Hz

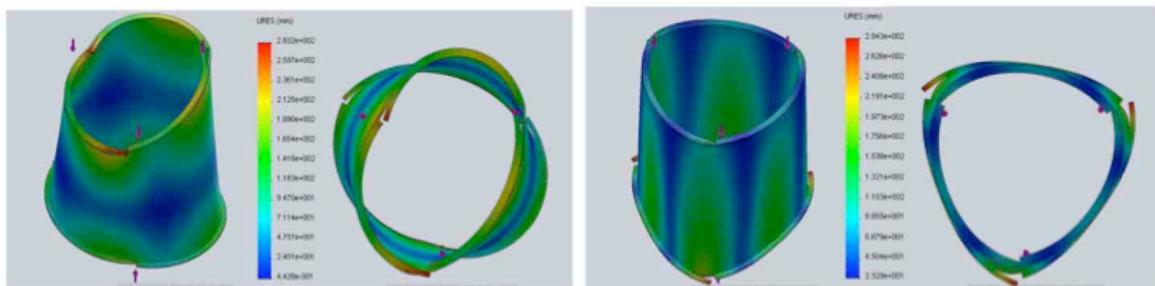

b) 43.22 Hz          c) 72.53 Hz

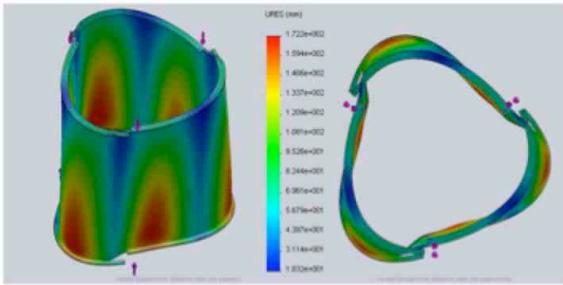

d) 88.01 Hz

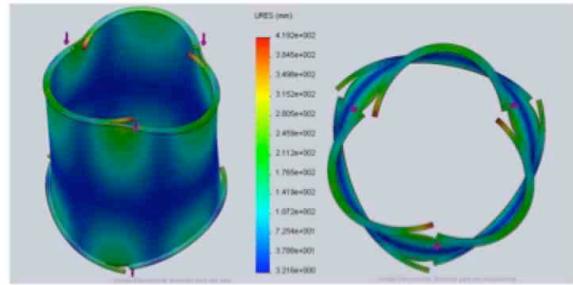

e) 110.29 Hz

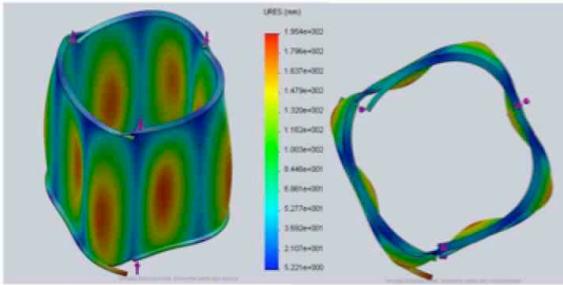

f) 127.02 Hz

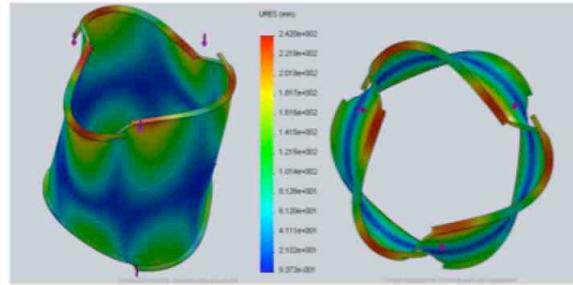

g) 142.77 Hz

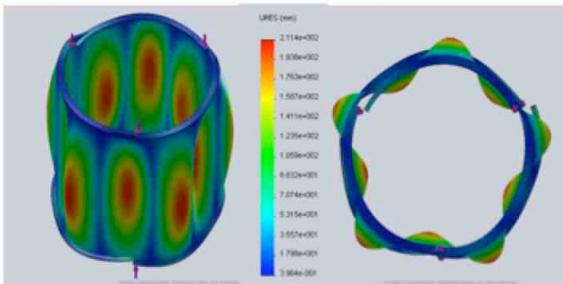

h) 148.85 Hz

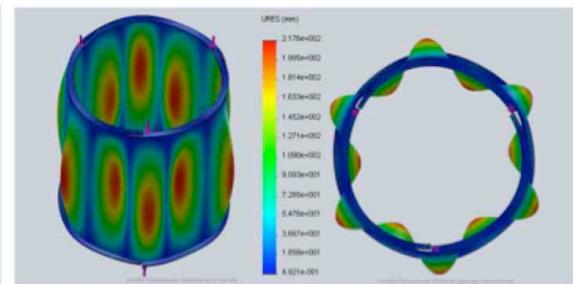

i) 182.31 Hz

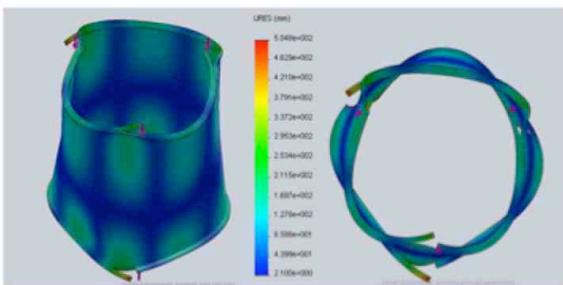

j) 221.92 Hz

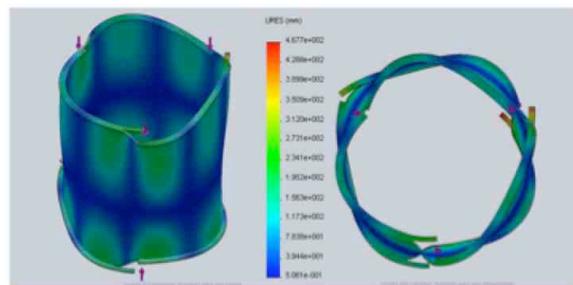

k) 224.18 Hz

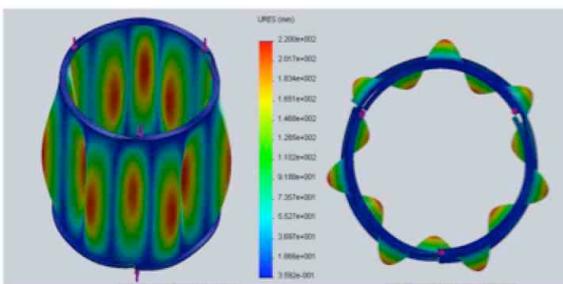

l) 235.06 Hz

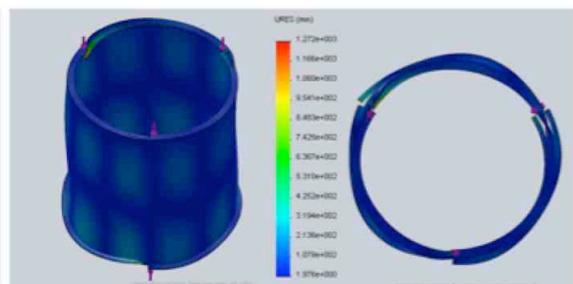

m) 282.04 Hz

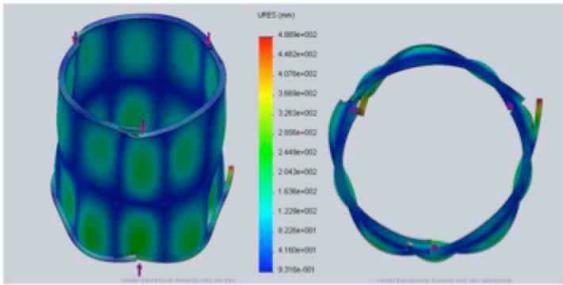
n) 284.24 Hz

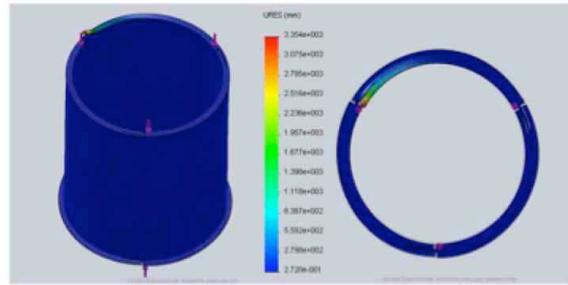
o) 292.54 Hz

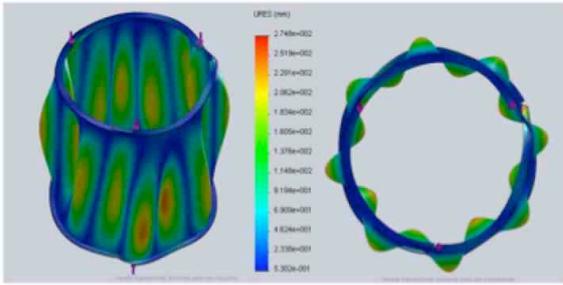
p) 301.20 Hz

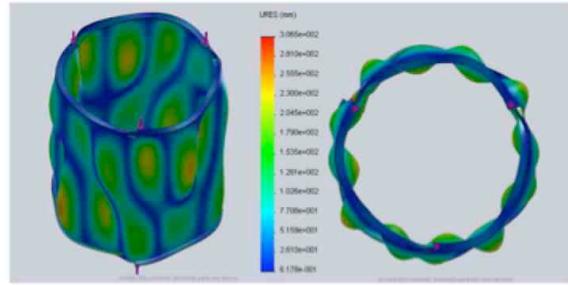
q) 303.03 Hz

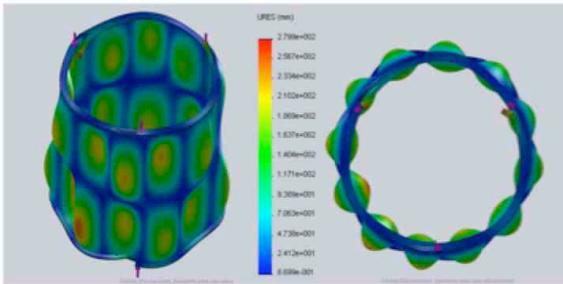
r) 307.70 Hz

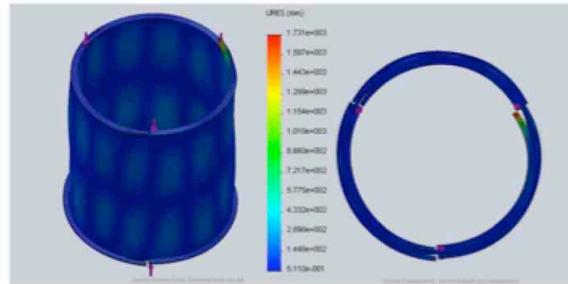
s) 329.15 Hz

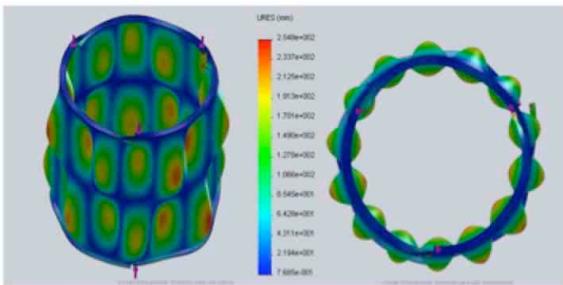
t) 331.11 Hz

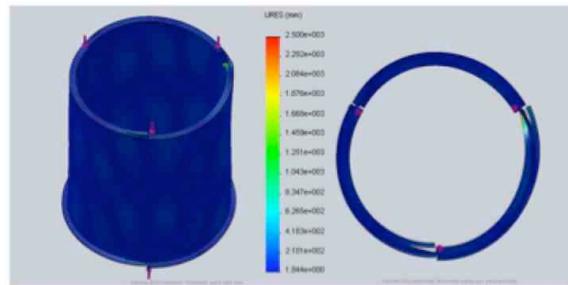
u) 334.22 Hz

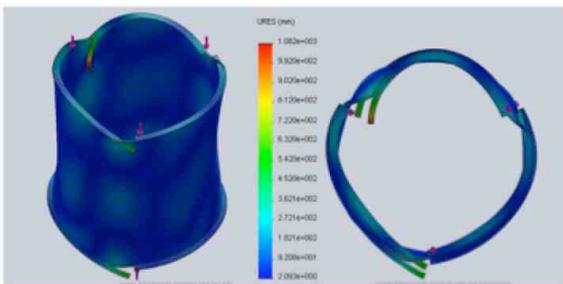
v) 341.54 Hz

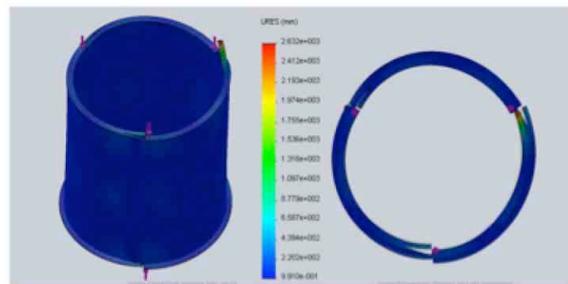
w) 350.24 Hz

FIG. 8. Many axial and/or radial-circumferential flexure modes simulated by SolidWorks[TM]. The first fundamental mode (i=2, j=0) in the top left hand corner is around 29 Hz, which is in agreement with the measured value.

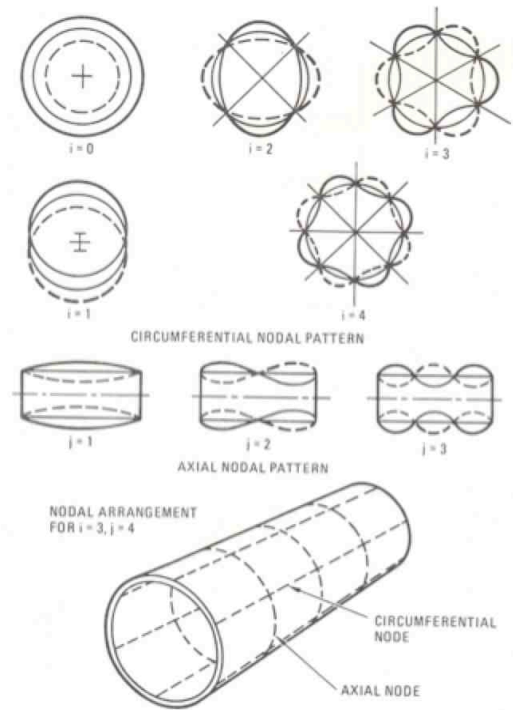

FIG. 9. Mode shape of axial and radial-circumferential flexure modes are shown[7].

In conclusion, with the exception of a measured mode at 6.6 Hz, which may be an external noise or a system vertical oscillation mode, all predicted and measured resonances are outside the 2-10 Hz target bandwidth. If this 6.6 Hz is a vertical oscillation mode, it will probably fall to below 2 Hz when the flexure arm flanges are used. For the other frequencies above 10 Hz, the present existing vibration isolation system of aLIGO will filter them out.

We should remember that this MNP system would not substitute the present existing vibration isolation system of aLIGO, but rather be added to it. There would be vibration isolation systems before this set of cylindrical shells of an evolved version for aLIGO, which would be HEPI and the active isolation platform. So, some seismic noise at these wire and cylindrical shell resonant frequencies would have already been suppressed. Also, all other cylindrical shells should also help to filter the low frequency resonances produced at these first stages. The resonances of the inner cylindrical shells would be perhaps more important, but again the initial noise reaching these last stages coming from the lab would be even lower.

## VII. FUTURE STUDIES

There are two major aspects for this multi-nested pendula system that we plan to investigate. Firstly, to discover how the cross coupling amongst the various degrees of freedom will affect the system. Exploring the present configuration, with high horizontal vibration isolation and poor vertical vibration isolation will help us to start this research. It will be completed later when we connect the rods to the long arm flange springs or replace them with gas springs using servo-controlled pressure to try to restore stability for tilting while at the same time, increasing the vertical vibration isolation. The second issue is to calculate the overall vibration isolation and cooling performance targets for an evolved prototype.

As previously mentioned before, the present architecture of these cylindrical shells is ideal for creating a cryogenic environment around the mirrors we wish to cool down. The innermost cylindrical shell can also be closed at the bottom and top with the exception of a few necessary holes, which include the one for the laser beam to reach the mirror. One important question to be studied is which cylindrical shell should be thermally grounded to a low temperature reservoir. Thermally grounding the innermost one has the advantage of placing the cryogenic reservoir close to the payload, which would make the cooling down of the mirror easier, but on the other hand all vibration produced by the cryogenic liquid boil-off would be introduced without the vibration isolation the multi-nested system could offer. If instead we thermally ground the first or second outermost shell, we will have to solve the problem of thermal conduction between the cylindrical shells. Right now, stainless steel rods, which have low thermal conductivity, are being used. We would have to consider substituting them with a high thermal conductivity material that has a high Q mechanical factor at cryogenic temperatures, such as silicon or sapphire fibers (silica fibers do not have high mechanical Q factors at low temperatures).

From an experimental point of view, in order to investigate the first issue, we will need to construct a large vacuum chamber to completely enclose the prototype and build a device to introduce a frequency variable controlled force (probably magnetic) on the outermost cylindrical shell. For the second issue, we will need to introduce intermediate vacuum chambers with super insulation sheets (made of aluminizing mylar).

## VIII. FINAL REMARKS

The resonant frequency of a pendulum is $\omega_o = (g/L)^{1/2}$ and a multistage vibration isolation has an ultimate $(\omega_o/\omega)^{2N}$ rolloff[1] if we keep the mechanical Q factor high. Therefore, the resonant frequency of each pendulum of an ideal multistage common equal pendula filter is $N^{1/2}$ times the resonant frequency of each pendulum of an ideal multistage *nested pendula* filter and, consequently, the latter will be better than the former by a factor of $N^N$, making this idea of a *nested pendula* a very interesting one.

However, in vibration isolation systems we have to make a compromise between thermal noise and vibration isolation noise. The former is minimized if we keep the mechanical Q high, but the latter needs a mechanical Q factor which is not so high, because the low frequency pendulum and vertical modes of the multi-nested pendula must be dampened somehow. On the other hand, if the mechanical Q decreases too much, the ultimate roll-off falls not as a second power of frequency per stage, but as a single power. These questions must be properly addressed in the future work mentioned in the previous section.

## ACKNOWLEDGMENTS

We acknowledge useful discussions with members of the LIGO Scientific Collaboration, in particular Warren W. Johnson, Harald Luck, Norna Robertson, Fabrice Matichard, Brian Lantz and other members of the Suspensions and Isolation Working Group (SWG), some of whom gave us important suggestions. We would also like to thank Riccardo deSalvo for useful discussions as well. ODA also would like to thank FAPESP (process # 2006/56041-3), CNPq and MCT/INPE for support. MCJr would like to thank CAPES for support.